\documentclass[usegraphicx]{mn2e}

\title{Non-emission line young
stars of intermediate mass}
\author[P. Manoj,~~G. Maheswar ~and~H. C. Bhatt.]
       {P.~ Manoj,~~G.~ Maheswar ~and~~H. C. ~Bhatt \\
        Indian Institute of Astrophysics, Bangalore 560034, India}
\date{}

\pubyear{2001}

\begin{document}

\maketitle
\begin{abstract}
        We present optical spectra of four intermediate mass candidate young
stellar objects that have often been
classified as Herbig Ae/Be stars. Typical Herbig
Ae/Be emission features are not present in the spectra of these stars.
Three of them, HD 36917, HD 36982 and HD 37062 are members
of the young Orion Nebula Cluster (ONC). This association
constrains their ages to be  $\le$ 1 
Myr. The lack of
appreciable near infrared excess in them suggests the absence of
hot dust close to the central star. But they do possess
significant amounts of cold and extended dust component  as revealed by the large
excess emission observed at far infrared wavelengths.
Fractional infrared luminosity $(L_{ir}/L_{\star})$ and the dust
masses computed from IRAS fluxes are systematically lower than
that found for Herbig Ae/Be stars but higher than those for
Vega-like stars. These stars may then represent the youngest
examples of Vega phenomenon known so far. In contrast, the other
star in our sample, HD 58647, is more likely to be a classical Be star as
evident from the low $L_{ir}/L_{\star}$, scarcity of circumstellar
dust, low polarization, presence of H$\alpha$ emission and near infrared excess 
and far infrared spectral energy distribution consistent with free-free emission
similar to other well known classical Be stars. 

\end{abstract}


\section{Introduction}
 Pre-main sequence stars of intermediate mass ($2 \leq M/M_{\odot} \leq 8$)
which show emission lines in their spectra 
 are called Herbig Ae/Be (HAEBE) stars. They were first
discussed as a group by Herbig (1960). He identified 26 stars which are  of spectral type A or B, located in obscured starforming regions and  illuminating a bright nebulosity in its immediate vicinity. Additions to the original
list were made by Finkenzeller \& Mundt (1984) and Herbig \&
Bell (1988). Recently, a more extended catalogue of HAEBE stars
and related objects was published by Th\'{e} et al. (1994). In
this catalogue 287 HAEBE stars and related objects are listed in
five tables, which include stars with later spectral type (G0 or
earlier) and those found relatively isolated from star forming
clouds. Out of this, only 109 stars which are listed in Table 1
of the catalogue, are recognized as either HAEBE stars or
potential candidate members. Other stars listed in the catalogue
are either of very uncertain or unknown spectral type or have not
been identified to belong to any specific group.

                      The following set of properties are often
taken as a working definition of HAEBE stars (Waters \& Waelkens 1998).  {\it (a) Spectral
type B or A with emission lines. (b) Infrared excess
due to hot or cool circumstellar dust or both. (c) Luminosity
class III to V}.  The emission lines are believed to be formed in a stellar wind originating
from a hot and extended chromosphere around the star
(Bouret \& Catala 1998), and is closely connected with the accretion related
disc activity. The near and far infrared excesses which
characterize the spectral energy distribution (SED) of these stars
is attributed to the presence of significant amounts of
circumstellar dust around them with a wide range in temperature
(e.g. Hillenbrand et al. 1992). Submillimeter and millimeter observations 
have clearly established  the existence of dust. Dust masses estimated
from these studies range from  $\sim 10^{-5}$  to $10^{-1} M_{\odot}$
(Waters \& Waelkens 1998 and references therein). Scattering of central star
light by the circumstellar dust manifests in 
large values of intrinsic polarization measured for HAEBE stars (
e.g. Grinin 1994).
However, the geometry of the circumstellar environment of HAEBE
stars is still a
matter of debate. Evidence for the presence of discs as well as
envelopes has been found. Recently Natta et al. (2001) have argued
that irradiated discs with a puffed-up inner wall of optically
thick dust provide a good fit to the observations over the entire
range of wavelengths.

                    Though the pre-main sequence (PMS) nature of the
HAEBE stars is now well established, several questions concerning
 their PMS evolution remain to be answered. Do all PMS stars
of intermediate mass go through the Herbig Ae/Be phase? What
happens to the attendant circumstellar material around a HAEBE star 
by the time it evolves into a main sequence star? There have been
suggestions in literature that HAEBE stars evolve into Vega-like
stars ( eg. Malfait et al. 1998, Waters \& Waelkens 1998).
Vega-like stars are characterized by substantial far infrared
excess due to cool dust, relatively low near infrared excess, low polarization and lack of emission lines in their spectra. The dust masses found around them are a few orders of magnitudes lower than that of HAEBEs. Also, Vega-like discs in general are  gas depleted (Lagrange et al. 2000). Do all HAEBE stars
pass through a Vega-like phase with gas depleted discs? These questions are critical to our understanding of the nature of the pre-main sequence evolution of intermediate mass stars. 

                   A study of non-emission line young stellar candidate objects
listed in Table 5 of the catalogue by  Th\'{e} et al. (1994) may
shed some light on the issues raised above. There are fourteen stars
listed in this table. Typical PMS
properties are less clearly seen in these stars. One of them, 
$\beta$ Pic, is a bonafide Vega-like star. The evolutionary status 
of other stars is not very clear. They are believed to be, as
the authors suggest, transition objects between PMS and MS phase.

            In this paper we present the results of a 
study of four non-emission line stars listed in the afore-mentioned catalogue.
Spectroscopic and polarimetric observations of
these stars were carried out. In Section 2 of this paper we present our
observations. Together with the
information  available from literature 
in different wavelength ranges, we discuss the structure of
circumstellar environment and the evolutionary status of these
stars in Section 3. Summary of  our
study is presented in Section 4.

\begin{table}
\centering
\begin{minipage}{75mm}
\caption{Log of spectroscopic observations}
\begin{tabular}{|l|l|c|}   \hline
 
Object    & Date of &  Exposure Time \\
          & Observation&               \\
\hline \hline
AB Aur    &  23 December 2000 &600s              \\
HD 36917  &  27 February 2002 & 600s             \\
HD 36982  &27 February 2002  & 600s               \\
HD 37062  &26 February 2002 & 300s               \\
HD 58647  & 22 December 2000 & 600s                \\
\hline
\end{tabular}
\end{minipage}
\end{table}

\section[]{Observations}
 Medium resolution ($ \lambda / \Delta
\lambda
\sim 3000$) optical CCD spectra were obtained   for
stars HD 36917 (V372 Ori), HD 36982 (LP Ori) and HD
37062 (V361 Ori) with the Optometrics Research (OMR) spectrograph on the 2.3 meter Vainu Bappu Telescope (VBT) and for HD 58647 with the Universal Astronomical
Grating Spectrograph (UAGS) on the 1 meter telescope at the Vainu
Bappu
Observatory, Kavalur, India.
 Log of spectroscopic observations
is given in Table 1. The prototype Herbig Ae/Be star $AB Aur$ was
also observed with the UAGS on the 1 meter telescope and is included in Table 1. All spectra were bias subtracted, flat-field corrected, extracted and wavelength calibrated in the standard manner using the IRAF{\footnote{  IRAF is distributed by National Optical Astronomy Observatories, USA.}} reduction package. In view of the presence of surrounding diffuse H$II$ region nebulosity in the direction of three of the programme stars, the background sky subtraction was performed in the following way. The stellar spectra were extracted by summing up over ten pixels (plate scale = $0.2\arcsec/pixel$) perpendicular to the dispersion axis. The sky background subtracted was obtained by summing up ten pixels, ten pixels ($2\arcsec$) away from the star on either side. The spectra were corrected for instrumental response and brought to a relative flux scale using the spectrophotometric standard observed on the same night.
Each spectrum spans a  wavelength range of $\sim 2400 \AA$, centered
roughly  at $H\alpha (\lambda \sim 6562 \AA)$.  Reduced spectra of HD 36917, HD 36982 and HD 37062  along with that of $AB Aur$ are presented in Figure 1(a) and that of HD 58647 in Figure 1(b).
                  Optical linear polarization
measurements were made with a fast
star-and-sky chopping polarimeter (Jain \& Srinivasulu 1991)
coupled at
the $f/13$ Cassegrain focus of the 1 meter telescope at the Vainu
Bappu
Observatory, Kavalur of the Indian Institute of Astrophysics. A
dry-ice
cooled R943-02 Hamamatsu photomultiplier tube was used as the
detector.
All measurements were made in the $V$ band with an aperture of $15\arcsec$
. The instrumental
polarization was determined by observing unpolarized standard stars  
from Serkowski (1974). It was found to be $\sim 0.1\%$, and has been
subtracted vectorially from the observed polarization of the programme
stars. The zero of the polarization position angle was determined by
observing the polarized standard stars from Hsu \& Breger (1982). The
position angle is measured from the celestial north, increasing
eastward. Observed polarization and position angle values are
presented in Table 2. Object name and date of observations are
presented in column 1 and 2 of the table. Percentage polarization
in V band and the probable error in the measurement are given in
columns 3 and 4 and position angle and probable error in columns 5 and 6.

\begin{table}
\centering
\begin{minipage}{75mm}
\caption{Polarization observations from Kavalur}
\begin{tabular}{|c|c|c|c|c|c|c|}   \hline 
Object    & Date of  & $P(\%)$ & $\epsilon_ {P(\%)}$
&$\theta(^{\circ})$&$\epsilon_{\theta}(^{\circ})$ \\
          & Observation      &          &                    &  &  \\
\hline \hline
HD 36982  & 12 March 1999& 0.53    & 0.11 &55& 8               \\
HD 58647  & 03 March 2000& 0.23    & 0.07 &123&7           \\
\hline
\end{tabular}
\end{minipage}
\end{table}

\begin{table}
\centering
 \begin{minipage}{75mm}
  \caption{Polarization measurements in $V$ band compiled from literature }
  \begin{tabular}{|c|c|c|c|c|c|c|} \hline
Object    & $P(\%)$ & $\epsilon_{ P(\%)}$
&$\theta(^{\circ})$&$\epsilon_{\theta}(^{\circ})$ & {Reference \footnote{
1. Heiles (2000), ~~~ 2. Oudmaijer et al. (2001)  }}\\
                        \\
\hline \hline
HD 36917  & 0.97& 0.032& 43& 0.9&1\\
HD 36982  & 1.01 & 0.021&56&0.6   & 1         \\
HD 37062  & 0.41 & 0.032&162&2.2& 1   \\
HD 58647  & 0.22&0.04&133.3&4.6   & 2           \\
\hline
   \end{tabular}
\end{minipage}
\end{table}

\begin{figure}
\resizebox{\columnwidth}{!}{\includegraphics*{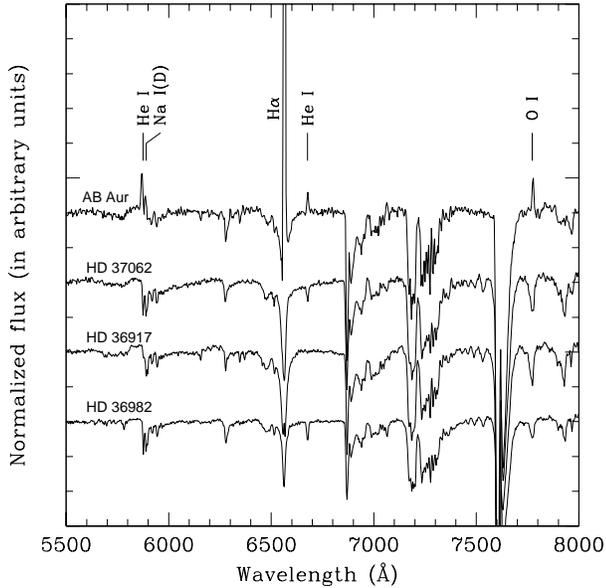}}
\caption{{\bf (a)}~~~Optical spectra of the Orion Nebula Cluster (ONC) member stars and AB Aur. In HD 37062, HD 36917 and HD 36982 typical Herbig Ae/Be emission lines are absent }
\end{figure} 

\setcounter{figure}{0}
\begin{figure}
\resizebox{\columnwidth}{!}{\includegraphics*{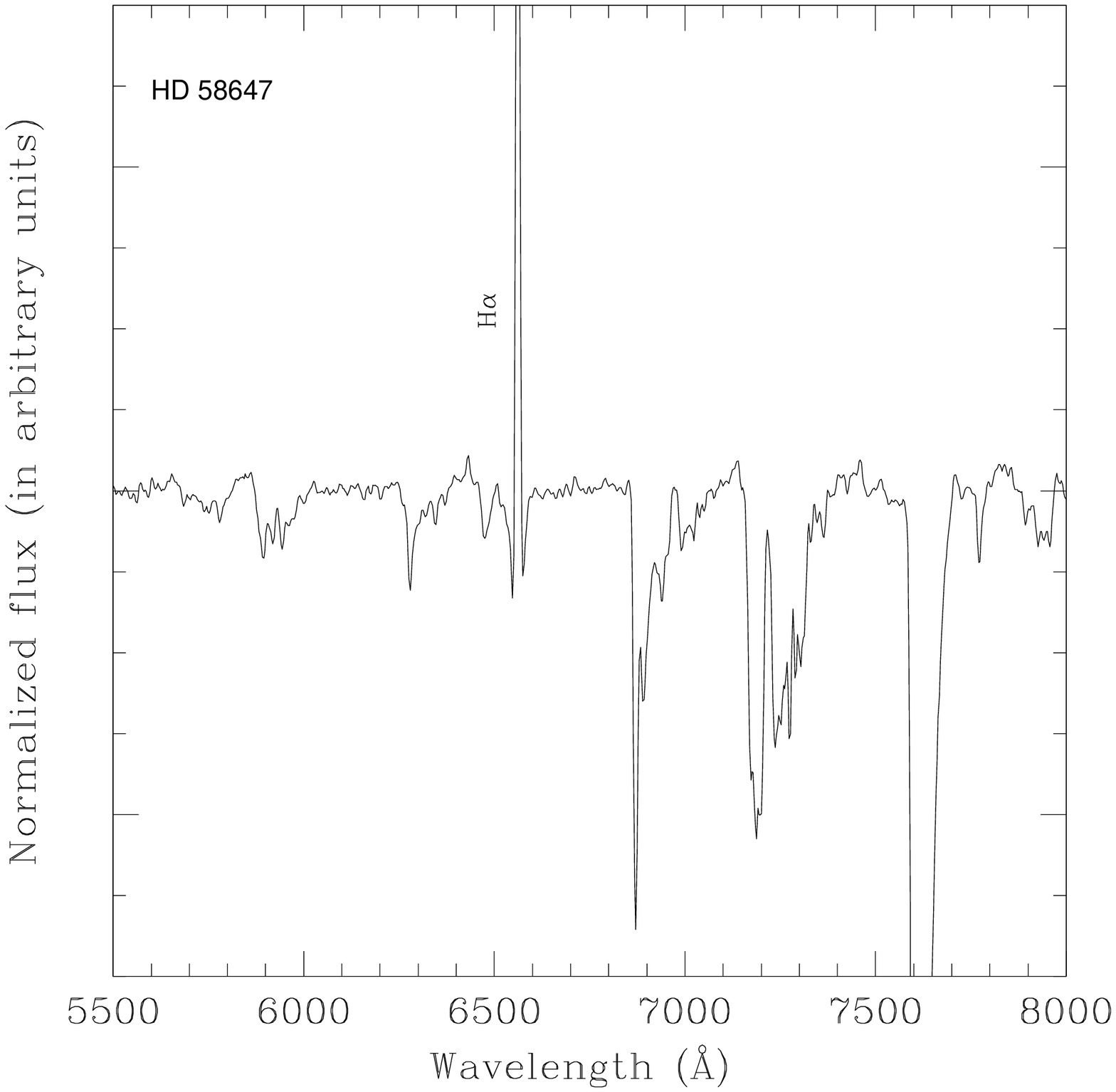}}
\caption{ {\bf (b)}~~~~Optical spectrum of HD 58647. Except $H\alpha$ no other Herbig Ae/Be emission lines are present}
\end{figure}  

\begin{figure}
\resizebox{\columnwidth}{!}{\includegraphics*{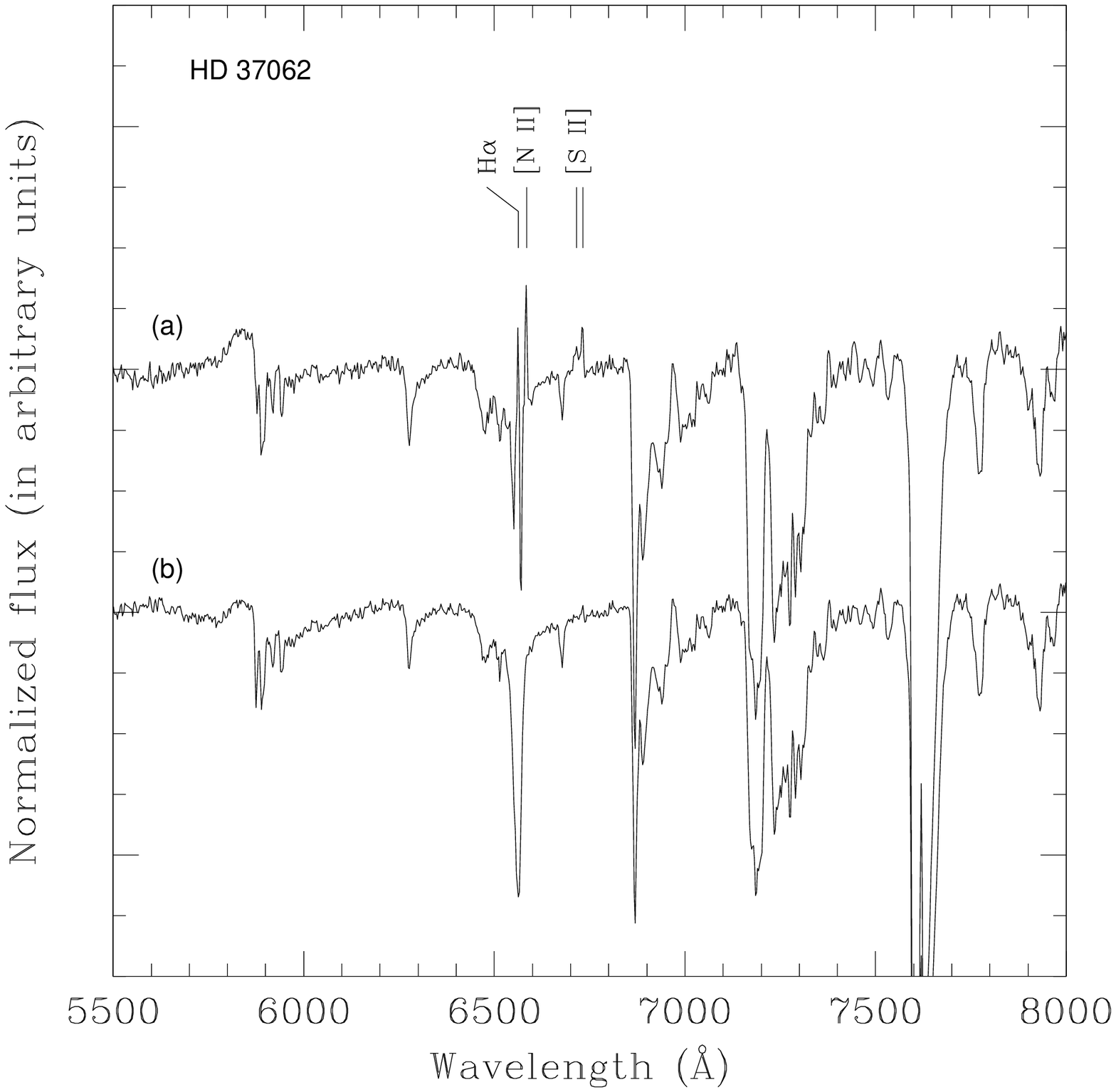}}
\caption{~~ (a) The raw spectrum of HD 37062 with nebular lines superposed on it ~~~ (b) The reduced spectrum after subtraction of the surrounding nebulosity }
\end{figure}

 It can be seen from Figure 1(a) and 1(b) that emission lines
are not present in the spectra of these stars except HD 58647. In
 Figure 1(a) we have also included a spectrum of $AB Aur$, 
a prototype Herbig Ae star,  in the same wavelength range and of similar resolution as
of the other spectra for comparison. Typical HAEBE
emission  features such as $H \alpha$, HeI
($\lambda5875$  \&  $\lambda 6678$) and OI ($\lambda 7774$)which are
prominent in $AB Aur$ are not seen in the
spectra of HD 36917, HD 36982 and HD 37062. In contrast, HD 58647 has a strong $H \alpha$ line in emission but
  does not show any other characteristic HAEBE
emission features. These stars cannot be unequivocally classified as
HAEBE stars. 
 
	We note that three of the programme stars, {\it viz.,} HD 36917, HD 36982,
HD 37062 are towards the direction of the Orion Nebula and the diffuse H$II$ region present there is projected on to the line of sight to these stars. Therefore a careful subtraction of the surrounding nebular emission from the observed spectra is very important. This is illustrated in Figure 2 which presents the raw spectrum with superposed nebular lines and the reduced spectrum with surrounding nebular emission subtracted for HD 37062.

  In Table 3 we present polarization data for these
stars already existing in literature. The polarization values measured
are typical of that of pre-main sequence stars except, again, for
HD 58647 which shows a relatively low value of polarization.
Also, LP Ori for which we have more than one polarization
measurements shows variability which is again an indicator of the
youth of the star.

\section{Discussion}

   All the stars in our sample have earlier been treated as
HAEBE stars  by different authors (e.g. Malfait et al. 1998, Yudin 2000, Valenti et al. 2000)
though in the catalogue by Th\'{e} et al. (1994) they are identified as
non-emission line stars. Spectra presented in Figure 1 show that the characteristic HAEBE emission features ($HeI$
$\lambda5875$  \&  $\lambda 6678$ and $ OI$ $\lambda 7774$) are
 conspicuously absent in  these stars.  Their classification as HAEBE stars is,
therefore, doubtful. A strong $H \alpha$ emission is present in HD 58647, but as argued later in this paper this star is likely to be a classical Be star. Weak $H \alpha$ emission, filling in the absorption core is present in HD 36917 even upon the subtraction of the possible contribution of the surrounding nebulosity. By comparing with synthetic spectra (Pickles 1985) of normal spectrophotometric standard stars of similar spectral type, we determine the equivalent width of the $H \alpha$ emission in this star to be $2.5$ $\AA$. This upper limit (for any residual nebular emission) to the $H \alpha$ equivalent width for HD 36917 is much smaller than those for typical Herbig Ae/Be stars (for the Herbig Ae/Be prototype star AB Aur we find $H \alpha$ emission equivalent width to be $44$  $\AA$). Interpreted as a measure of accretion rate (e.g. Hillenbrand et al. 1992, Muzerolle et al. 2001) this $H \alpha$ emission line flux implies mass accretion rate ($\dot{M}$) of order $\sim 10^{-7}$
which is  much smaller ( by a factor of 10 to 100) than those in typical Herbig Ae/Be stars ($\dot{M} \sim 10^{-6}M_{\odot}/yr - 10^{-4}M_{\odot}/yr$ depending on the spectral type; Hillenbrand et al. 1992).      
                     
                    Three of the stars that we have observed {\it
viz.}, HD 36917, HD 36982 and HD 37062 are in the direction of the
known H$II$ region (Sharpless No. 281)  towards the the Orion Nebula. 
From proper motion and radial velocity studies it has
been established that these stars are kinematically connected
with the Orion Nebula Cluster (ONC) which is at the northwestern
end of Orion A cloud (van Altena et al. 1988, McNamara \& Huels
1983, McNamara 1976, Tian et al. 1996, Hillenbrand 1997). They
belong to the stellar population in the inner $\sim 2.5$ $pc$ of
the ONC which is at a distance of $470\pm 70$ $pc$ (Genzel et al. 1981; Walker 1969; Tian et al. 1996; Hillenbrand 1997). ONC as a whole is characterized by a mean age of $<1Myr$ and an age spread which is probably less than 2 Myr with an internal velocity dispersion of $\sim 2$ $kms^{-1}$ (Hillenbrand 1997; Tian et al. 1996). The kinematic
association with the Orion Nebula Cluster strongly  constrains the ages of these
 stars to within a few $\sim 10^6$ years. They are
young stars of intermediate mass. In the following we use the
average distance to the ONC $(470$ $ pc)$ as the distance to its member stars.

\begin{table*}
\begin{center}
\begin{minipage}{120mm}
\caption{Data compiled from literature and quantities estimated
from them}

\begin{tabular}{|l|l|c|c|c|c|c|c|c|c|}   \hline 
Object    & Sp. Type &$vsini$    & B & V & J & H & K & $E(B-V)$ \\
          &          &$kms^{-1}$ &mag&mag&mag&mag&mag&mag  \\
\hline \hline

{HD 36917\footnote{Spectroscopic binary (Lavato \& Abt 1976)}}& B9.5+A0.5  &117  & 8.18 & 8.06 &7.269&7.012&6.615& 0.14 \\
HD 36982& B1.5V& 98  & 8.48 & 8.44 &7.754&7.762&7.521& 0.29 \\
HD 37062& B4V  & 78  & 7.80 & 8.24 &7.850&7.725&7.576& -0.26 \\
HD 58647& B9IVe & 280 & 6.82 & 6.85 &6.464&6.115&5.433& 0.10  \\
\hline
\end{tabular}
\end{minipage}
\end{center}
\end{table*}

\begin{figure}
\resizebox{\columnwidth}{!}{\includegraphics*[20mm,130mm][150mm,230mm]{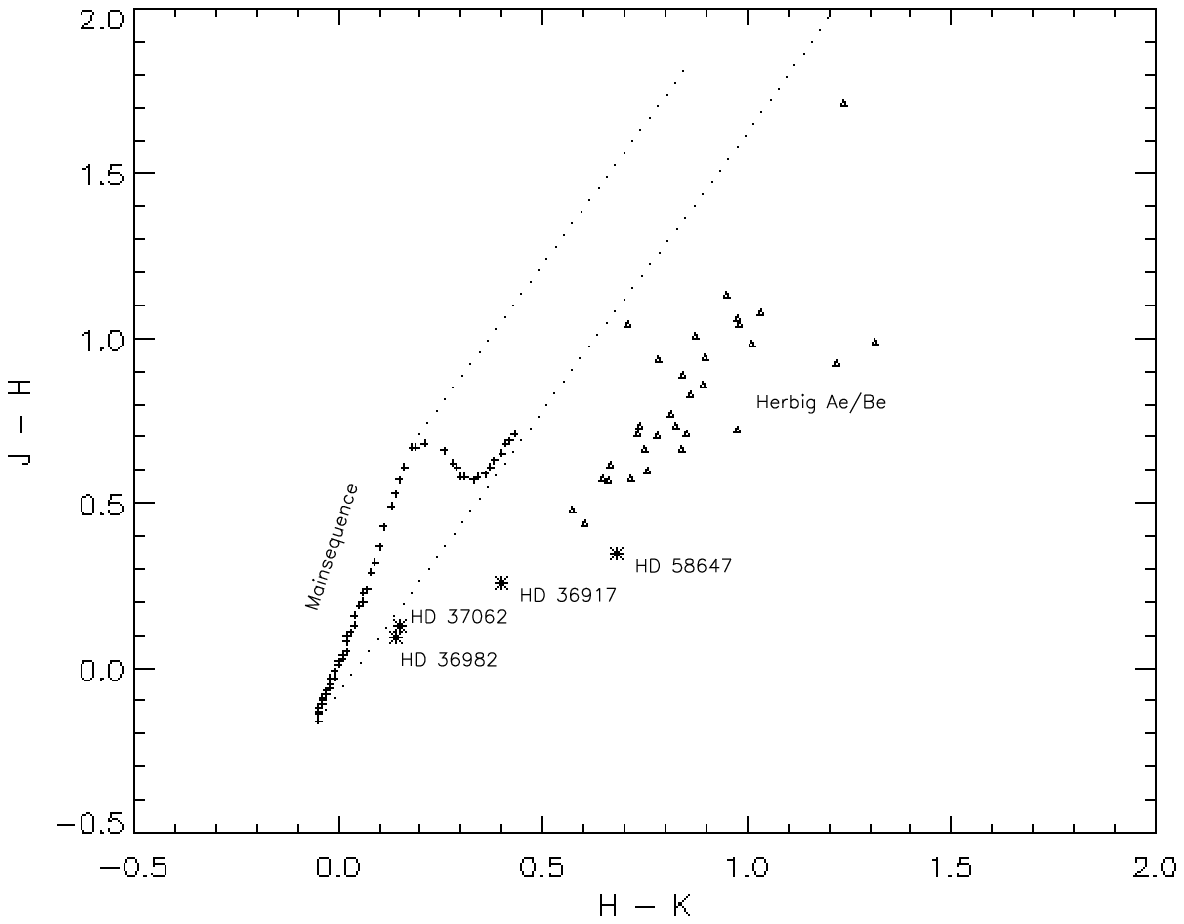}}
\caption{Near infrared colour-colour diagram for the stars. Also
plotted are Herbig Ae/Be stars (triangles) and main sequence stars
 (crosses).}
\end{figure} 
 
         To complement our observations, we have
collated available information from literature on our sample
stars. We present them in Table 4.  The spectral types are taken
from SIMBAD except that of HD 36917 for which it is from Lavato \& Abt (1976). $vsini$ values are taken form Yudin (2000) and van den Ancker et al. (1998). The optical B and V magnitudes are derived from Tycho
magnitudes except for HD 37062 for which B and V magnitudes are from SIMBAD. Tycho photometric errors are less than 0.02 mag. The J, H, K magnitudes are from 2MASS
catalogue. Errors in J,H,K magnitudes are less than 0.07 mag.
Reddening,$E(B-V)$ estimated from spectral types and photometric magnitudes  is listed in column 9. 

                We have constructed near infrared (NIR) colour-colour 
       diagram from the 2MASS magnitudes
 for our programme stars which is shown in Figure 3.  Along with
the four programme stars, HAEBE stars and main sequence stars are also
plotted in the diagram. The colours for the main sequence stars are
from Koornneef (1983). HAEBE stars are taken from Th\'{e} et al.
(1994) and their colours are derived from 2MASS magnitudes. The
two parallel dotted lines form reddening band for normal stellar
photospheres. These lines are parallel to the reddening vector and
bound the range in the colour-colour diagram within which stars
with purely reddened normal stellar photospheres can fall (Lada \& Adams 1992). It can be seen that all the four stars
are distinctly separated from the region occupied by HAEBE stars.
Their near infrared charcterestics are different from that of
HAEBE stars. The near infrared excesses of all the four stars are
considerably lower than that of HAEBE stars.  HD 36982 and HD
37062 have very little near infrared excess, if any. The near
infrared
excess in HAEBE stars is attributed to  reradiation from hot dust
less than  $\sim 1AU$ from the star. The low near infrared excess shown by our programme stars, then,  
 would strongly suggest the absence of submicron sized dust grains close to the star.  \\

In Table 5 we present far infrared data for the stars and the
quantities estimated from them except for HD 37062 which does not
have an IRAS entry. HD number of the stars, their IRAS source names and the IRAS flux densities at 12, 25, 60 and 100$\mu$ are listed in the first six columns. In column 7 we list the fractional infrared luminosity ($L_{ir}/L_{\star}$) estimated from IRAS fluxes. $L_{ir} =
4 \pi d^2 F_{ir}$ with  \\\\ $F_{ir} =
[20.653f_{12}+7.53f_{25}+4.578f_{60}+1.762f_{100}]\times 10^{-14}
Wm^{-2}$\\ \\ and $d$ being the distance to the star (Cox 2000). The IRAS flux
densities (Jy)  at 12, 25, 60 and 100 microns are given by
$f_{12}$, $f_{25}$, $f_{60}$, and $f_{100}$ respectively.
$L_{\star}$ is computed from $M_v$ using standard bolometric corrections (Cox 2000) where $M_v$ is evaluated from dereddened $V$ magnitude and distance to the star. Since the star HD 36982 is located below the ZAMS in the colour - magnitude diagram (cf. Fig. 4) $L_{\star}$ for this star is taken to be the ZAMS luminosity expected for it's spectral type.  The black body colour temperatures derived from the ratio of fluxes at
$25\mu$ and $60\mu$ are considered to be the dust temperatures and are listed in column 8. The dust temperature derived for HD 36917 is a lower limit since the IRAS $60\mu$ flux density is only an upper limit. Dust masses (in units of $M_{\oplus}$, the mass of earth = $6\times10^{27}g$) listed in column 9 are computed using the relation,
\begin{center}   $ M_d = 4 \pi a \rho_d d^2 F_{ir}/3Q_{\nu} \sigma T_d^4 $ 
\end{center}
assuming a grain size of $a=1\mu$, dust grain material density $\rho_d =2g/cc$, absorption efficiency $Q_{\nu}=0.5$, and with $F_{ir}$ computed from the flux densities at four IRAS bands. Since IRAS flux densities at $100\mu$ are upper limits for all the stars we have used flux densities expected at $100\mu$  for the derived colour temperatures in estimating $F_{ir}$.

 The fractional infrared luminosities $L_{ir}/L_{\star}$  estimated
for HD 36917 and HD 36982 are quite significant ($L_{ir}/L_{\star}$ $\sim 0.15$). Here it is assumed that the IRAS flux densities quoted do represent emission from these sources and are not due to other sources in the IRAS beam. We note that IRAS point sources 05323-0536, 05327-0529 and 07236-1404  have positional coincidences with HD 36917 , HD 36982 and HD 58647 to within $4\arcsec$, $2\arcsec$ and $2\arcsec$ respectively. The surface density of IRAS point sources in this region is $\sim 3\times10^{-3}$ to $10^{-2} sources/(arcmin)^2$. Therefore, for a given star, in the 12 and 25 $\mu$ bands, the probability that an unrelated IRAS point source is in the IRAS beam ($0.75\arcmin \times 4.5\arcmin$) is only $\sim 1-3\%$. If the IRAS fluxes observed for these stars are dominated by thermal emission from the circumstellar dust present around these stars, which is most likely the case in view of the relatively low surface density of IRAS point sources and low probability of chance projections, then for the derived dust temperatures listed in Table 5 the emitting dust is at distances of $\sim 10 AU$ (HD 36917) and $\sim 100AU$ (HD 36982) from the central star, assuming the dust to be distributed in an optically thick disc as in some models of HAEBE stars (Hillenbrand et al. 1992). On the other hand if the dust is distributed in optically thin shells as in Vega-like stars, then the dust is located at $\sim 100 AU$ (HD 36917) and $\sim 1000 AU$ (HD 36982) from the stars. These dimensions of dust shells or rings are quite similar to those of Vega-like stars and are compatible with the dust responsible for emission being part of the circumstellar environment. It is clear from Table 5 that these two stars have significant amounts of cold circumstellar dust present around them. The dust masses estimated are systematically lower than those for HAEBE stars but higher than that found around Vega-like stars.

                         In Figure 4 we present the far infrared
spectral energy distribution for the three stars. Their IRAS flux
densities are plotted against the wavelength
together with the expected photospheric flux densities at IRAS
wavelengths (Song 2001) for a star of that spectral type. It is
clearly seen that HD 36917 and HD 36982 have considerable excess
at far infrared wavelengths  whereas  HD 58647 has a low
excess and an energy distribution that is decreasing with
wavelength.  

                      It is clear from the above discussion that
the three member stars of Orion Nebula Cluster (ONC) are
extremely young and lack emission lines in their spectra. This
would imply the absence of a hot and extended chromosphere around
them where the emission lines are thought to be formed. In the case of HD 36982 and HD 37062, which are early B
type stars it is possible that this is an evolutionary effect.
They are possibly at the end of their pre-main sequence phase, which
is roughly $\sim 1-2$ $Myr$. The absence of $HeI$ line in emission,
which is an indicator of accretion, also suggests that these stars
are beyond their accretion phase. Almost complete absence of near
infrared excess further supports the absence of an inner accretion
disc. This evolutionary picture is further supported by the fact
that HD 36917, which is of a later spectral type and thus has a longer
PMS life time, shows near infrared
excess though at a much lower level than seen in HAEBEs. Binarity of the star cannot account for the near infrared excess. The $J-H$ and the $H-K$ colours computed by adding up the individual fluxes expected for each component of the binary ($B9.5+ A0.5$) and estimating the combined magnitudes, are only $0.006$ and $-0.002$ whereas the observed colours are 0.257 and 0.397 respectively. Thus it could be concluded that the
inner disc has not been completely disrupted in this star. . 

\begin{figure}
\resizebox{\columnwidth}{!}{\includegraphics{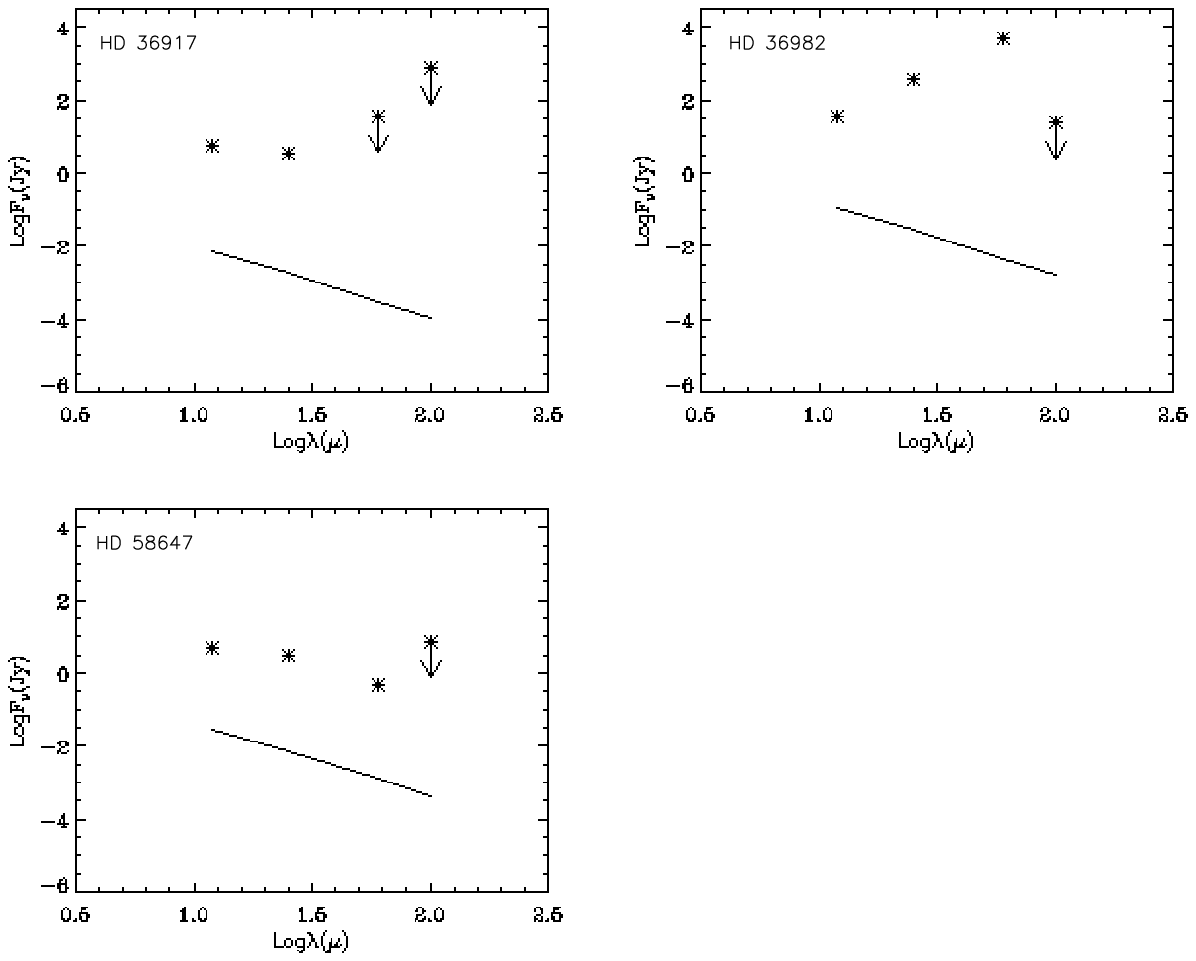}}
\caption{Far infrared spectral energy
distributions for the stars. The solid line represents the
photospheric flux density expected from the star at IRAS
wavelenghts. Asterisks are
IRAS flux densities measured for the stars. Arrows pointing downwards indicates
that flux density given is an upper limit.}
\end{figure}

         However, we do not rule out the possiblity of the stars
not passing through a HAEBE phase with emission lines and near
infrared
excess. A different formation mechanism or a very destructive
cluster environment can drastically alter the PMS properties and
evolutionary sequence that a young star passes through.

\begin{table*}
\begin{center}
\begin{minipage}{140mm}
\caption{Far infrared data for the stars and the quantities estimated from them}
\begin{tabular}{|l|l|l|l|l|l|c|c|c|}   \hline 
Object&  IRAS source&  \multicolumn{4}{c|}{IRAS Flux Densities, $f_{\nu}$ (Jy)}&
$L_{ir}/L_{\star}$&   $T_d$&    $M_d/M_{\oplus}$   \\ 
      & Name&$f_{12}$&$f_{25}$&$f_{60}$&$f_{100}$&  &(K)&  \\ \hline \hline
   
HD 36917  &05323-0536  &5.73&3.37&$37.8^*$&$740^*$ &0.13 &$\geq 70$  &$\leq 2$  \\ 
HD 36982  &05327-0529  &33.9&367&4800&$24.5^*${\footnotetext{* Upperlimits}}& 0.17 & 68  & 230  \\
HD 58647  &07236-1404  &4.95&2.87&0.47 &$7.36^*$&0.01  &$^{\dagger}10^4$&$^{\dagger}5\times10^{-10}${\footnotetext{$\dagger$ Unphysical, since the colour temperature is much larger than the dust sublimation temperature (see text)}}    \\
\hline
\end{tabular}
   
\end{minipage}
\end{center}
\end{table*}

                           In any case these stars fit well with
the definition of Vega-like stars though the far infrared
excesses, fractional infrared
luminosities $(L_{ir}/L_{\star})$, and the dust masses computed
for HD 36982 and HD 36917 are much higher than that for the
prototype Vega-like stars and for
"old PMS" (OPMS) and "young main sequence" (YMS) systems discussed
recently by Lagrange et al. (2000). Their circumstellar dust may not be
the debris product but rather what is left over from their PMS
phase. These stars, then, are the youngest
Vega-like stars hitherto known.

                      Unlike the stars which are members of the
Orion Nebula Cluster, HD 58647 may not be a young star. This object is about 30 degree away from the Orion complex and is not associated with any star forming cloud. Though it shows
an excess at the near infrared wavelengths, it's far infrared excess is small. $L_{ir}/L_*$  indicates the scarcity of dust present around the
star. Low polarization observed reinforces this fact. Colour temperature derived from the IRAS fluxes is very high and is much larger than the dust sublimation temperature. The dust
mass estimated formally from IRAS fluxes is negligibly low and is unphysical as the colour temperature does not represent the dust temperature. Clearly, the excess shown by the star in the IRAS wavelengths cannot be attributed to circumstellar dust. 
The infrared excess, both near and far, most likely is  due to free-free
emission from an ionized region around the star. We suggest that
HD 58647 is a classical Be star. The relatively large value of
$vsini \sim 280$ $kms^{-1}$  supports this claim. The presence of
$H\alpha$ line in emission observed in the spectra is consistent
with the star being a classical Be. Moreover, its IRAS flux
density distribution varies nearly as $F_{\nu}$ $ \alpha$
$\nu^{0.6}$ which is very similar to that for classical Be
stars (Taylor  1990). 

\begin{figure}
\resizebox{\columnwidth}{!}{\includegraphics{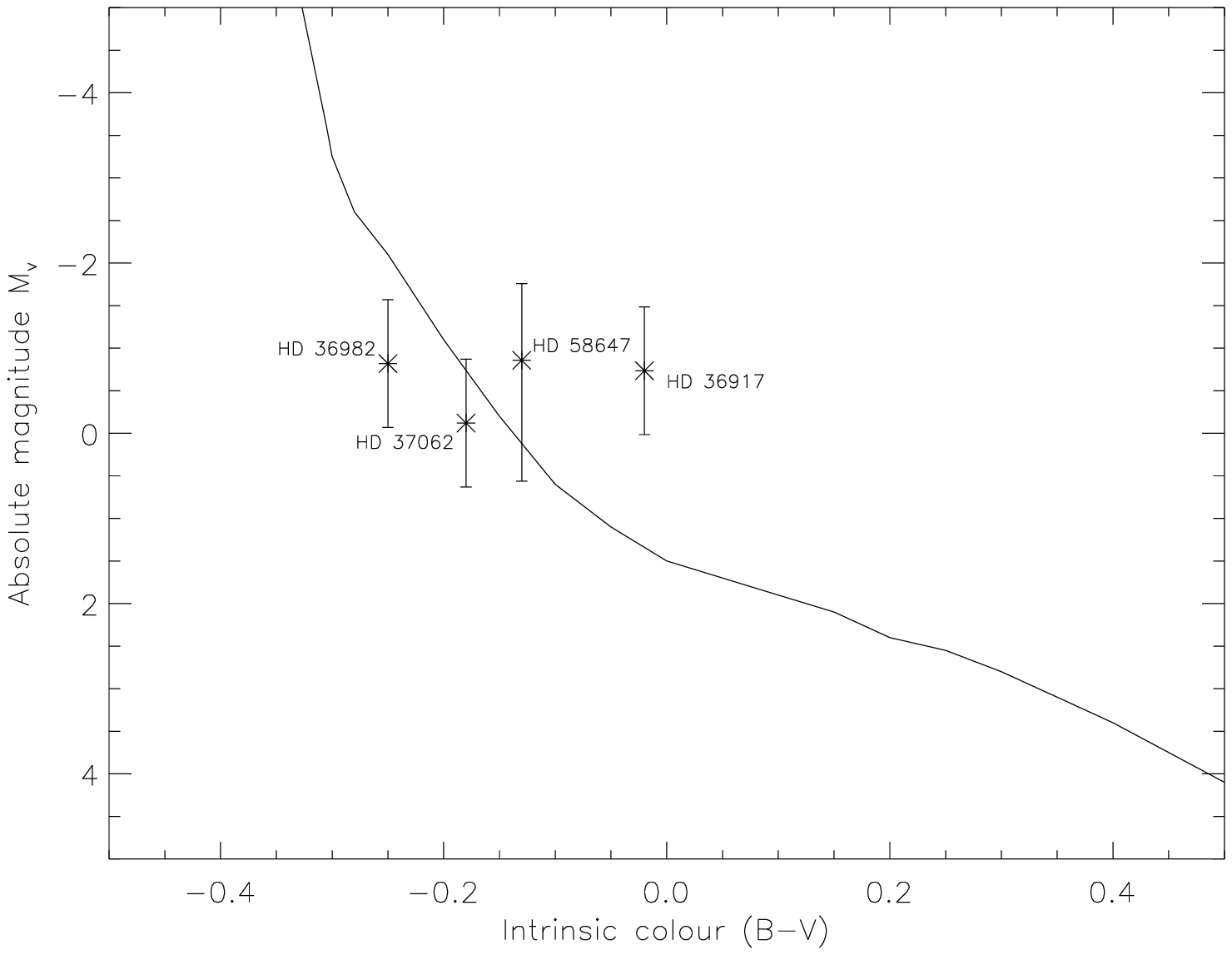}}
\caption{Colour-absolute magnitude diagram for the stars. The
solid line represents the zero-age main sequence.}
\end{figure}

                             A colour-absolute magnitude diagram
constructed for all the four programme stars is shown in Figure 5.
The zero-age main sequence data is taken from Schmidt-Kaler
(1965).  For the ONC member stars a distance of $470$ $pc$ is
assumed and a {\it Hipparcos} distance of $277$ $pc$ is used for HD 58647 in computing the absolute magnitude. Major contribution to the error bars shown in the figure results from uncertainities in distance. Further, the average interstellar value of 3.1 is used for the  ratio $R=A_v/E(B-V)$ . Significance of their position in the colour-magnitude diagram (CMD) is discussed below.

                HD 36917 which is a spectroscopic binary  (B9.5 + A0.5)  (Lavato \& Abt 1976) is
found to be way above the main sequence in the colour - magnitude diagram. Increase in the
brightness caused by binarity would only account for $\sim 0.7
mag$. Thus it's location in the colour - magnitude diagram is 
consistent with the star being a pre-main sequence and the
age indicated by its kinematic association with the ONC.             
               
                    The other two stars which are members of the ONC, {\it viz}., HD 36982 and HD 37062 fall below the zero-age
main sequence (in particular HD 36982) in the CMD. A possible explanation for their anomalous position is that they have an anomalous circumstellar extinction component. If they have already reached the zero age main sequence then the extinction towards HD 36982 and HD 37062, implied by their location in the CMD and computed from the observed and absolute V magnitudes for a distance of $470$ $pc$
are  $\sim 2.12$ $ mag$ and $\sim 0.7$ $ mag$ respectively. Such large extinction and a rather low  $E(B-V)$
cannot be produced by interstellar grains of submicron size. This
would imply the presence of large grains around these stars. The neutral extinction produced by these grains can be even larger than that estimated assuming the stars to be zero age main sequence stars, without causing any change in the colour excess. The possiblity of these stars being intrinsically brighter than a ZAMS star of similar spectral type cannot be ruled out. They could be pre-main sequence stars with very high extinction. \\

\section {Summary}

                We have obtained optical spectra of four
non-emission line stars listed in the catalogue by Th\'{e} et al.
(1994). Emission lines which characterize the typical HAEBE spectrum
are not seen in any of the stars. Also, their near infrared properties are very unlike HAEBE stars. The amount of circumstellar dust present is also systematically lower than that found around HAEBE stars. We argue that these stars
cannot be unambigu

ously classified as HAEBE stars. Nevertheless, three of the
stars which are kinematic members of Orion Nebula Cluster are  
very young. This association constrains their ages to be less
than a few $Myr$. Absence of emission lines, low near infrared excesses and presence of far infrared excesses in these objects make them somewhat similar to Vega-like stars though the dust masses estimated for them are much higher than that of prototype Vega-like stars. These stars may then represent the youngest examples of Vega phenomenon and may well be the
intermediate mass counter parts of weak line T Tauri stars. The
observed reddening and estimated extinction for these stars
indicate presence of larger than submicron sized dust grains around these
stars. One of the stars in our
sample, HD 58647, is more likely to be a classical Be star as
evident from the low $L_{ir}/L_*$, scarcity of circumstellar dust, low polarization, presence of $H\alpha$ emission and appreciable near infrared excess and far IR spectral energy distribution consistent with free-free emission similar to other well known classical Be stars. \\ \\

{\bf ACKNOWLEDGMENTS}\\ 
      We thank the referee Dr. M. van den Ancker for critical comments and useful suggestions.

\end{document}